\documentclass[prb,twocolumn,showpacs,groupedaddress]{revtex4}
\usepackage{amssymb}
\usepackage{amsfonts}
\usepackage{amsmath}
\usepackage{graphicx}

\begin{document}
\title{Density matrix renormalization group approach of the spin-boson model}
\author{Hang Wong}\email{frensel@gmail.com}
\author{Zhi-De Chen}\email{tzhidech@jnu.edu.cn}
\affiliation{Department of Physics, Jinan University, Guangzhou
510632, China}

\begin{abstract}
We propose a density matrix renormalization group approach to tackle
a two-state system coupled to a bosonic bath with continuous
spectrum. In this approach, the optimized phonon scheme is applied
to several hundred phonon modes which are divided linearly among the
spectra. Although DMRG cannot resolve very small energy scales, the
delocalized-localized transition points of the two-state system are
extracted by the extrapolation of the flow diagram results. The
phase diagram is compared with the numerical renormalization group
results and shows good agreement in both Ohmic and sub-Ohmic cases.
\end{abstract}
 \maketitle

\section{Introduction}
The density matrix renormalization group (DMRG) is an important tool
for studying the strongly correlated systems in low
dimensions.\cite{White, Schollwoeck} In the past decade, one
significant limitation of DMRG---finite basis requirement in the
model which involves infinite degree of freedom, e.g., phonon
states, was circumvented by a controlled truncation
technique.\cite{Zhang} This technique is applied to many models,
typically, such as 1D Holstein model,\cite{Zhang} 1D
Holstein-Hubbard model,\cite{Weisse} spin Peierls
model,\cite{Friedman} and spin-boson model.\cite{Nishiyama} The main
idea of this technique, controlled truncation, is realized by the
density matrix approach which is useful for finding the most
probable states of the truncated system. By its light, the infinite
Hilbert space can be reduced to governable dimensions without
significant loss of accuracy. However, the truncation technique is
originally designed for the systems involving just one phonon mode,
i.e., the Einstein model. The direct application of this technique
to the spin-boson model with a continuous spectrum of phonon modes,
is not very successful.\cite{Nishiyama} For instance, in the case of
Ref.\ \onlinecite{Nishiyama}, the number of phonon modes were
limited to $N=18$, the physics of this highly discrete model may be
unreliable. Furthermore, the number of states of each phonon mode
kept is $m=2$, the truncation error is relatively large and no
convincing result on the delocalized-localized transition was
found.\cite{Nishiyama} These limitation implies that, to handle the
system with many phonon modes in a DMRG treatment, one needs to
develop an improved truncation technique. This is the motivation of
the present paper.

Here, let us briefly introduce the spin-boson model. The spin-boson
model is an important toy model in the study of dissipative quantum
systems. Its Hamiltonian is given by (set $\hbar=1$)\cite{Leggett,
Weiss}
\begin{equation}\label{eq1}
    H=\frac{\Delta}{2}\sigma_x+\frac{\epsilon}{2}\sigma_z+\sum_i\omega_ia_i^\dag
    a_i+\sigma_z\sum_i\lambda_i(a_i+a_i^\dag),
\end{equation}
where the Pauli matrices $\sigma_{x,z}$ describe a two-state
system, $a_i^\dag$ and $a_i$ are phonon creation and annihilation
operators with frequencies $\omega_i$ for the $i$-th phonon modes,
$\epsilon$ is an additional bias (asymmetry), $\Delta$ is the bare
tunneling splitting, and $\lambda_i$ represents the coupling
between the two-state system and the $i$-th phonon mode.
Generally, the so-called bath spectral function
\begin{equation}\label{eq2}
    J(\omega)=\pi\sum_i\lambda_i^2\delta(\omega-\omega_i)
\end{equation}
completely determine the solutions of the spin-boson model. With an
energy cutoff $\omega_c$, i.e., discards the high energy modes, the
bath spectral function has a power-law form
\begin{equation}\label{eq3}
    J(\omega)=\frac{\pi}{2}\alpha\omega^s\omega_c^{1-s},
\end{equation}
where $\alpha$ is a dimensionless coupling constant which
characterizes the dissipation strength, $0<s<1,s=1,$ and $s>1$
represent sub-Ohmic, Ohmic, and super-Ohmic dissipation,
respectively. The primary purpose of the spin-boson model is to
study the effect of the environment on quantum tunneling of the
two-state system. Here the environment is modelled as a collection
of harmonic oscillators, which serves as the origin of
dissipation.\cite{Leggett, Weiss} Intuitively, the presence of the
environment will make the tunneling particle as a ``dressed" one,
just like the electron in a polaron-phonon (or exciton-phonon)
system, and therefore its quantum tunneling decreases as the
coupling increases. One important issue in spin-boson model is to
study the phonon-induced localization (also stated as
delocalized-localized transition), i.e., how  quantum tunneling dies
out as the coupling (or the dissipation)
increases.\cite{Leggett,Weiss,sp,keh,Silbey, Chin, Wong, Lv,
Bulla,Li}  Such a delocalized-localized transition at $T=0$ is now
considered as some kind of quantum phase transition called boundary
phase transition.\cite{qpt,qp} 

In general, the Hamiltonian of the spin-boson model cannot be solved
exactly, especially in the sub-Ohmic case.\cite{Weiss} The
delocalized-localized transition has been widely studied by various
methods with different approximations, yet a consensus on the
delocalized-localized transition in sub-Ohmic case is still lacking.
By integrating out the bath degrees of freedom, the spin-boson model
was mapped to an Ising model and the localized transition was
predicted to exist for $s\le1$ (i.e., in both Ohmic and sub-Ohmic
cases).\cite{sp,keh} However, the path-integration with the
so-called noninteracting blip approximation (NIAP) and the adiabatic
renormalization predicted that no delocalized-localized transition
happens in the sub-Ohmic case.\cite{Leggett} On the other hand, in
the sub-Ohmic case, variational calculations, flow equation method,
and other perturbation calculations predicted a discontinuous
delocalized-localized transition,\cite{Chin, Wong, keh, Lv} while
the non-perturbative numerical renormalization group (NRG)
calculation shows a continuous one.\cite{Bulla} Recently, the
authors showed that the discontinuous transition in the sub-Ohmic
case obtained by the variational calculation is simply an artifact
of the variational scheme due to the fact that the energy of the
variational ground state can no longer be lower than the energy of
the trial ground state (displaced-oscillator state).\cite{Chen}
While this result sheds some lights on the problem, the discrepancy
between various treatments mentioned above has not yet been
resolved. In addition, although the NRG approach is regarded as the
most powerful tool for treating the phase transition, the error due
to discretization in numeric calculation has to be
considered.\cite{Lv} Under this sense, the delocalized-localized
transition is necessary to study by another non-perturbative method,
i.e., DMRG. We hope that this paper will be helpful for resolving
the discrepancy.

The organization of this paper is as follows. In the following
section, we propose a finite system DMRG approach with controlled
truncation technique to the spin-boson model, a thousand phonon
modes can be treated. In Sec.\ \ref{sec3}, we determine the DMRG
parameters and discuss the very small energy scales limitation of
our treatment. Sec.\ \ref{sec4} suggests a extrapolation scheme to
circumvent the very small energy scales limitation. The
delocalized-localized transition points of the spin-boson model,
which is associated to the very small energy phonon modes, are
obtained by extrapolation of ``pseudo-critical" points. Conclusion
is given in the last section.

\section{The finite system DMRG algorithm of the spin-boson model}
Here we present a finite system DMRG algorithm with the optimized
truncation of multi-modes phonon space to treat the spin-boson model
whose bosonic bath involves several hundred phonon modes. The key
strategy of the algorithm is that one represents a single phonon
mode as a site. The spin-boson model therefore becomes a finite-size
chain. In this case, finite system DMRG is naturally applied to this
model since it is appropriate to reduce the environment error with
sweeping processes.\cite{Legeza} To reach this, we must divide the
frequency spectrum into $N$ intervals, i.e.,  $[\nu_{i-1}, \nu_i]$,
where $i=1,\ldots,N$, $\nu_i-\nu_{i-1}=\nu_{i+1}-\nu_i$, $\nu_0$=0,
$\nu_N=\omega_c=1$, and $\omega_i=(\nu_i+\nu_{i-1})/2$. In other
words, the frequency spectrum is divided linearly. The corresponding
coupling parameters $\lambda_i$ can be obtained by the spectral
function (\ref{eq2}) and (\ref{eq3})
\begin{equation}\label{eq4}
    \lambda_i^2=\frac{1}{\pi}\int_{\nu_{i-1}}^{\nu_i}J(\omega)d\omega=
    \frac{\alpha\omega_c^{1-s}}{2(s+1)}(\nu_i^{s+1}-\nu_{i-1}^{s+1}).
\end{equation}
Suppose that $N_b$ bare phonon states
($|0\rangle,|1\rangle,\ldots,|N_b-1\rangle$) are sufficient to
represent one phonon mode accurately, therefore we can limit $N_b$
bare phonon states in each phonon site. Using the controlled
reduction technique,\cite{Zhang} the dimension of each phonon mode
can be further reduced to $m$ where $m<N_b$. However, even though
$m=2$ is quite large for a dozen phonon modes, as in the treatment
by Nishiyama.\cite{Nishiyama} In this case, the number of phonon
modes that can be treated is seriously restricted. Our solution to
this problem is to truncate a set of phonon sites with the density
matrix approach within each DMRG step, i.e.,  we do not optimized
phonon modes individually. The truncated multi-phonon sites can be
continuously optimized by the sweeping of the finite system DMRG
algorithm. Similar treatment was done by Friedman in the study of
spin-Peierls model.\cite{Friedman} With these prerequisites, the
finite system DMRG algorithm can be implemented in the following
way.

\begin{figure}[h]
  \includegraphics[width=0.5\textwidth]{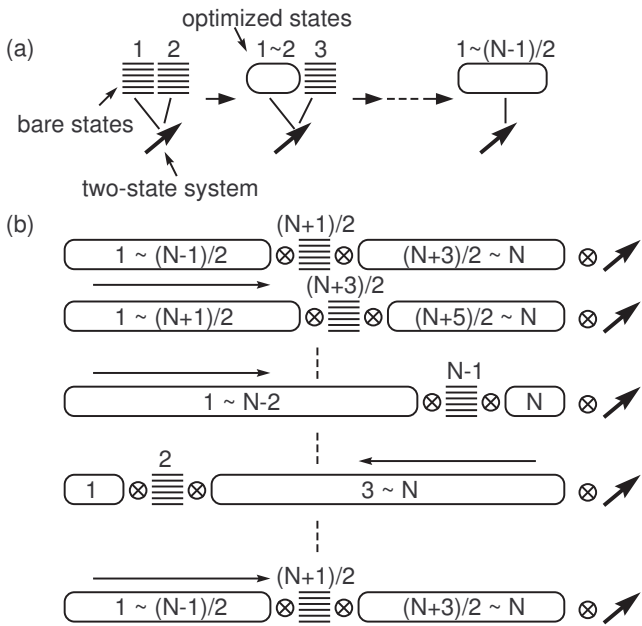}\\
  \caption{(a) the warmup procedure of the finite system DRMG
  algorithm, where the numbers represent the phonon modes.
  This figure shows the warmup procedure of the left part phonon
  modes $1\sim (N-1)/2$; see Fig. \ref{fig1}(b). The right part
  phonon modes can be obtained by similar fashion.
   (b) Systematic illustration of the finite system DMRG algorithm. This figure shows
   one sweep in the algorithm.}\label{fig1}
\end{figure}

As the standard finite system DMRG algorithm which is used in
Heisenberg model,\cite{White} the first step of the algorithm is
``warmup". We must generate a series of phonon blocks for the
subsequent sweeping processes of the finite system DMRG algorithm.
For simplicity, we assume the number of phonon modes is odd and
generate the blocks $1\sim(N-1)/2$ and $(N+3)/2\sim N$ separately,
where different numbers represent different phonon modes. With this
simplification, all the phonon blocks
$1\sim2,1\sim3,\ldots,1\sim(N-1)/2,(N+3)/2\sim N,(N+5)/2\sim
N,\ldots,N-1\sim N$ except for the blocks $1$ and $N$ which keep
$N_b$ bare phonon states are limited to a $2\times2$ matrix because
only the two-state system have been traced out;\cite{note1} see
Fig.\ \ref{fig1}(a).
During the course of warmup, each phonon mode with $N_b$ bare phonon
states is added to the preceding block. After a truncation with
density matrix approach, one new block is generated. We shall show
that $N_b=10$ is sufficient for the implementation of our algorithm
in most cases. Note that, every block generated within the warmup
processes must be stored in memory for later use.

Secondly, the finite system DMRG algorithm is implemented as in
Fig.\ \ref{fig1}(b). The finite system DMRG algorithm is more or
less the same as the standard algorithm.\cite{White} The main
difference between the two algorithms is that we add one site
within each DMRG step instead of two sites. It is because there
are no interactions between the phonon blocks, the implementation
of our algorithm is identical to the standard algorithm, and no
further correction is needed.\cite{White2} For convenience, the
two-state system can be placed on leftmost side or rightmost side
on the chain, as to calculate the reduced density matrix and apply
the traditional wave function transformations technique to the
finite system algorithm.\cite{White3} Within each DMRG step, a
phonon mode with $N_b$ bare states is added. This new site is used
to generate a new phonon block or optimize the old phonon block
with $M$ optimized states. Finally, the energies of the target
states will converge after one or two sweeps are preformed.

In summary, the algorithm can be proceeded as follows:
\begin{enumerate}
  \item warmup, generating a series of phonon blocks for subsequent
  sweeping processes;
  \item starting at the center phonon mode $(N+1)/2$, adding one phonon
  mode with $N_b$ bare states to the chain;
  \item performing the sweeping process to the whole chain;
  \item if the energies of the target states are not
  converged after a sweeping, then return to step (2).
\end{enumerate}

\section{Discussion and the limitation of the algorithm}\label{sec3}
One important issue of the algorithm is that how to choose the
parameters $N,N_b,M$, and the number of sweeps $N_s$. Unlike the
spin $1/2$ Heisenberg chain, there are no interactions between the
phonon blocks, the number of states kept per block is not quite
large. Therefore, it may be possible to treat a thousand phonon
modes while the number of states kept per block is never needed more
than $M=20\sim30$. In general, there are only $7-8$ largest
eigenvalues in the reduced density matrix of the phonon block have
significant values.
\begin{figure}[h]
  \includegraphics[width=0.5\textwidth]{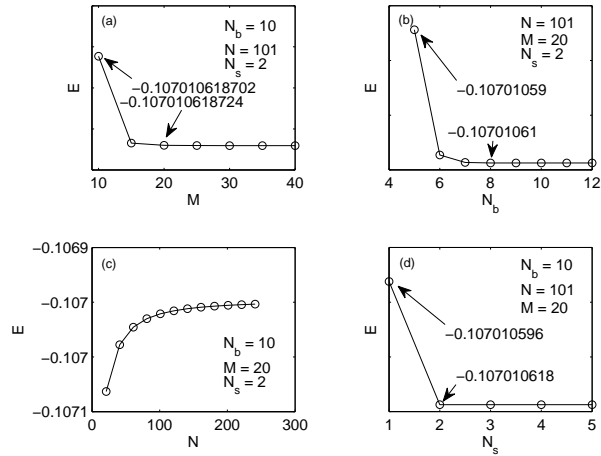}\\
  \caption{Dependence of the ground state energy on the parameters $N,N_b,M$,
   and $N_s$ for $\epsilon=0,s=0.6,\alpha=0.1$, and $\Delta=0.1$. (a) dependence on $M$ for
   fixed $N,N_b$, and $N_s$; (b) dependence on $N_b$ for fixed $N,M$, and $N_s$; (c) dependence on the number of
   phonon modes $N$ for fixed $N_b,M$, and $N_s$; (d) dependence
   on $N_s$, the number of sweeps versus the ground state energy, for fixed $N_b,N$, and $M$.}\label{fig2}
\end{figure}
The dependence of the ground state energy on the parameters
$N,N_b,M$, and $N_s$ for $s=0.6,\alpha=0.1$, and $\Delta=0.1$ is
shown in Fig.\ \ref{fig2} (targeting the ground state only, but
the following conclusions are also true for targeting both the
ground state and the first excited state). It is worth noting that
even though $M=10,N_b=6$, and $N_s=1$ can give rather the same
results. However, the number of phonon modes $N$ will highly
affect the results. This is also the main limitation of our DMRG
strategy.

\begin{figure}[h]
  \includegraphics[width=0.3\textwidth]{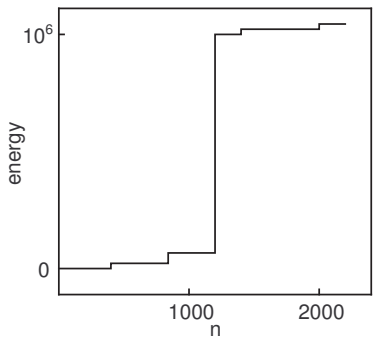}\\
  \caption{This figure is a schematic scaled energy spectrum of the DMRG or NRG calculation
  when one deals with the very small energy phonon modes. Here $n$ is the number of energy levels.
  The vertical coordinate (energy) is scaled by $N$ and $\Lambda^N$ in DMRG and NRG, respectively.}\label{fig3}
\end{figure}

As indicated in Refs.\ \onlinecite{Nishiyama}, \onlinecite{Bulla},
and \onlinecite{Bulla2}, the very small energy phonon modes are
important for revealing the critical phenomena, e.g., the
delocalized-localized transition of the two-state system. However,
the strategy of our algorithm needs linear discretization of the
spectrum which can not resolve very small energy scale. If one tries
to apply a logarithmic discretization which is used in NRG to the
DMRG algorithm, the energy levels of the Hamiltonian emerge a
staircase-like aspect when one deals with the very small energy
phonon modes; see Fig.\ \ref{fig3}. The DMRG scheme fails in this
situation because the target states cannot be determined. This
difficultly stems from the truncation strategy of the DMRG, say, it
iteratively calculates the lowest eigenstates for finding the most
probable states of the decimated system. However, in practice, it is
harsh for the iterative diagonalization routines being used by DRMG,
such as Lanczos and Davidson, to converge when the staircase-like
energy levels occur.
In fact, the staircase-like energy levels also appear in NRG
calculations. Nevertheless, the truncation scheme of NRG, which
retains the lowest-lying states directly, is simply to implement
in this situation. In other words, the performance of the standard
diagonlization routine used by NRG will not be affected by the
``shape" of the spectrum while DMRG needs iterative diagonlization
routine which converges arduously.

DMRG cannot resolve very small energy scales, this limitation is
serious. It implies that the critical coupling $\alpha_c$ cannot be
determined due to the energy levels cannot reach to a fixed point
without very small energy phonons\cite{Bulla} and the spin-spin
correlation function cannot be calculated in very small energy
scales. Furthermore, the effective tunneling splitting
$\Delta_r=\langle\sigma_x\rangle$ also is not adequate to identify
the critical coupling $\alpha_c$ because it fails to characterize
the tunneling in equilibrium in the sub-Ohmic case.\cite{Anders} Our
DMRG calculations have the same conclusion, namely, $\Delta_r\neq0$
when $\alpha>\alpha_c$ in the sub-Ohmic case and
$\Delta_r\rightarrow0$ when $\alpha\rightarrow\alpha_c$ (note that
$\alpha_c$ is a function of $s$ and $\Delta$) in the Ohmic case (not
presented). Moreover, the entanglement entropy method proposed by
Ref. \onlinecite{Hur} recently, which is used to determine the
critical couplings and performs very well in NRG, is not working as
expected in DMRG when the very small energy information is lacking.

\section{Extracting the critical points by
extrapolation}\label{sec4} 
\begin{figure}[h]
  \includegraphics[width=0.45\textwidth]{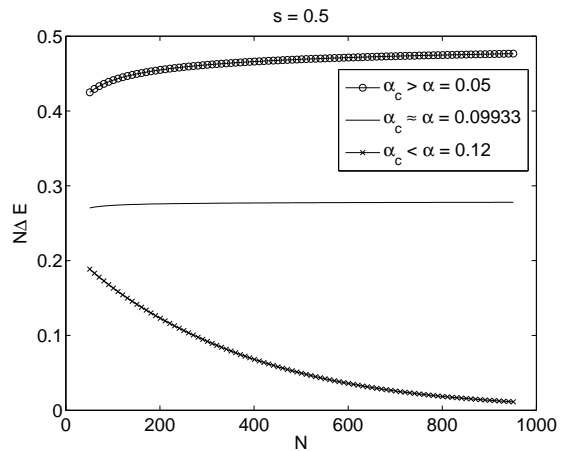}\\
  \caption{Dependence of the scaled energy different between ground state
  and excited state $N\Delta E$ on the number of phonon modes $N$ for $\alpha<\alpha_c$,
  $\alpha\approx\alpha_c$, and $\alpha>\alpha_c$. Parameters are $\epsilon=0,s=0.5,N_b=10,M=20$, and $N_s=2$.}\label{fig4}
\end{figure}

Now, we seek to show that the critical
couplings $\alpha_c$ can be determined by extrapolating the
pseudo-critical couplings $\alpha_c'$ which are extracted in a
``DMRG flow" to thermodynamic limit.\cite{Boschi} Similar to the
energies in logarithmic discretization of NRG which are falling off
as $\Lambda^{-N}$,\cite{Bulla} the energies are falling off as
$N^{-1}$ in linear discretization. Therefore, one can target the
ground state and the first excited state and scale the energy gap
$\Delta E=E_{\mathrm{excited}}-E_{\mathrm{ground}}$ as $N\Delta E$
and plot the flow diagram $N\Delta E$ versus $N$. As one can see in
Fig.\ \ref{fig4}, the flows of $N\Delta E$ are qualitatively
different within two regime $\alpha<\alpha_c$ and $\alpha>\alpha_c$.
Therefore, we assume that there exist a function $\alpha_c'(N)$
which separates the two regimes, where $N$ is relatively small in
comparison with thermodynamic limit. In practice, $\alpha_c'(N)$ can
be easily determined by a bisection process of two couplings
$\alpha<\alpha_c'(N)$ and $\alpha>\alpha_c'(N)$ with the slope of a
line segment consist of the scaled energies of two points
$[N-1,N+1]$. We conceive that the pseudo-critical coupling
$\alpha_c'(N)$ will converge to the critical coupling $\alpha_c$
when $N\rightarrow\infty$ since the fixed points are reached.

Accordingly, the extrapolation of the $\alpha_c'(N)$ versus $1/N$
curve determines the critical coupling $\alpha_c$ at the limit of
$1/N\rightarrow0$.
\begin{figure}[h]
  \includegraphics[width=0.45\textwidth]{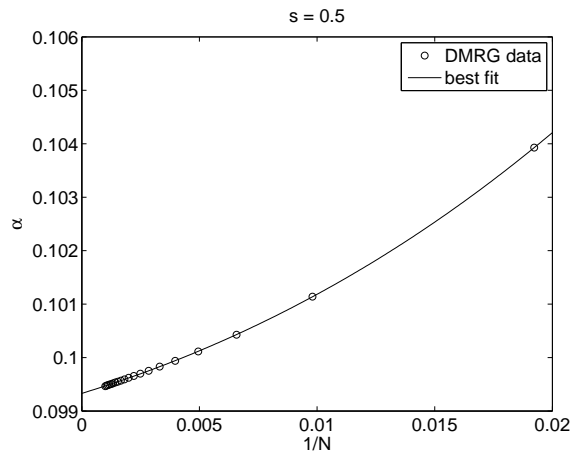}\\
  \caption{Dependence of the pseudo-critical couplings $\alpha_c'(N)$(circle)
  and the best-fit values(line) on the inverse number of phonon modes $1/N$. Parameters are
$\epsilon=0,s=0.5,N_b=10,M=20$, and $N_s=2$.}\label{fig5}
\end{figure}
Figure \ref{fig5} shows the best-fit of the pseudo-critical
couplings $\alpha_c'(N)$ with $s=0.5$. At $1/N=0$, it turns out that
$\alpha_c\approx0.09933$. In fact, the best-fit curves
$\alpha_c'(1/N)$ are somewhat different for sub-Ohmic and Ohmic
dissipation. It is related to the fact that the transition in the
sub-Ohmic case is characterized by a quantum critical fixed point in
contrast to the Ohmic case.\cite{Bulla} Since we cannot find a
formula to fit all the cases, the extrapolations are done by the
simplest polynomial fitting.

Admittedly, it might be doubted that if $N$ is small, $\alpha_c'$
could be inappropriate for extrapolation since it is inconsistent
with the $\alpha_c'$ which are obtained with large $N$. For
instance, there are some cases show that the infinite system DMRG is
a better choice to tackle this problem.\cite{Capone, Juo, Juo2}
However, on the one hand, our strategy of the DMRG in the spin-boson
model limits the implementation of the algorithm. In order to
``insert" the spin-boson model into the DMRG algorithm, one must cut
the spectrum of the bosonic bath to finite number of pieces and
therefore the spin-boson model becomes a finite-size chain. Before
performing the linear discretization, the number of sites $N$ and
the coupling constant $\lambda_i$ must be determined. Naturally, it
brings about the finite system DMRG algorithm to handle this model
and a ``real" infinite DMRG algorithm is difficult to implement in
practice. On the other hand, in our finite system DMRG solution, the
number of phonon modes treated is relatively large. We carefully
check the calculations and find that when we calculate the
$\alpha_c'$ with $1/N\le0.005$, the $\alpha_c'$ are always
monotonic. Hence, the extrapolations are safe and correct in our
treatment. Furthermore, references \onlinecite{Capone} and
\onlinecite{Juo} also performed an extrapolation of the number of
states kept, but the result shown in Fig.\ \ref{fig2} and the fact
of non-interacting phonon blocks guarantee that this quantity is not
significant in our calculation notwithstanding.

\begin{figure*}
    \centering
    \begin{minipage}[c]{0.45\textwidth}
        \centering
        \includegraphics[width=1\textwidth]{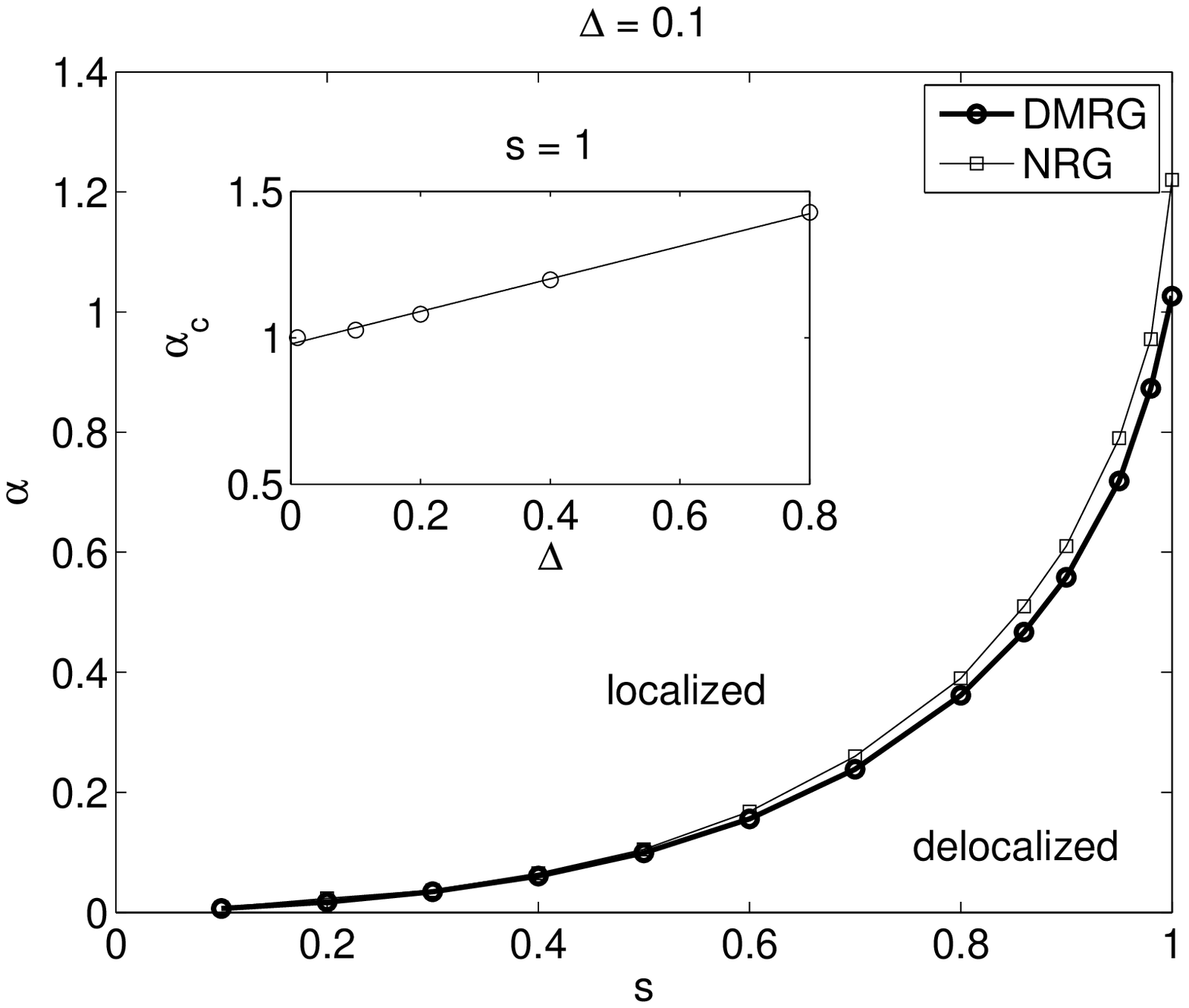}
        \caption{The delocalized-localized transition boundary of the spin-boson model.
        The DMRG data are compared with the NRG data (Ref.\ \onlinecite{Bulla}, PRL).
        Inset: dependence of the $\alpha_c$ on the parameter
        $\Delta$ for Ohmic case and the related linear fit $\alpha_c(\Delta)=0.56\Delta+0.98$.
        Parameters are $\epsilon=0,N_b=10,M=20$, and $N_s=2$.}\label{fig6}
    \end{minipage}
    \begin{minipage}[c]{0.05\textwidth}
        \mbox{}
    \end{minipage}
    \begin{minipage}[c]{0.45\textwidth}
        \centering
        \includegraphics[width=1\textwidth]{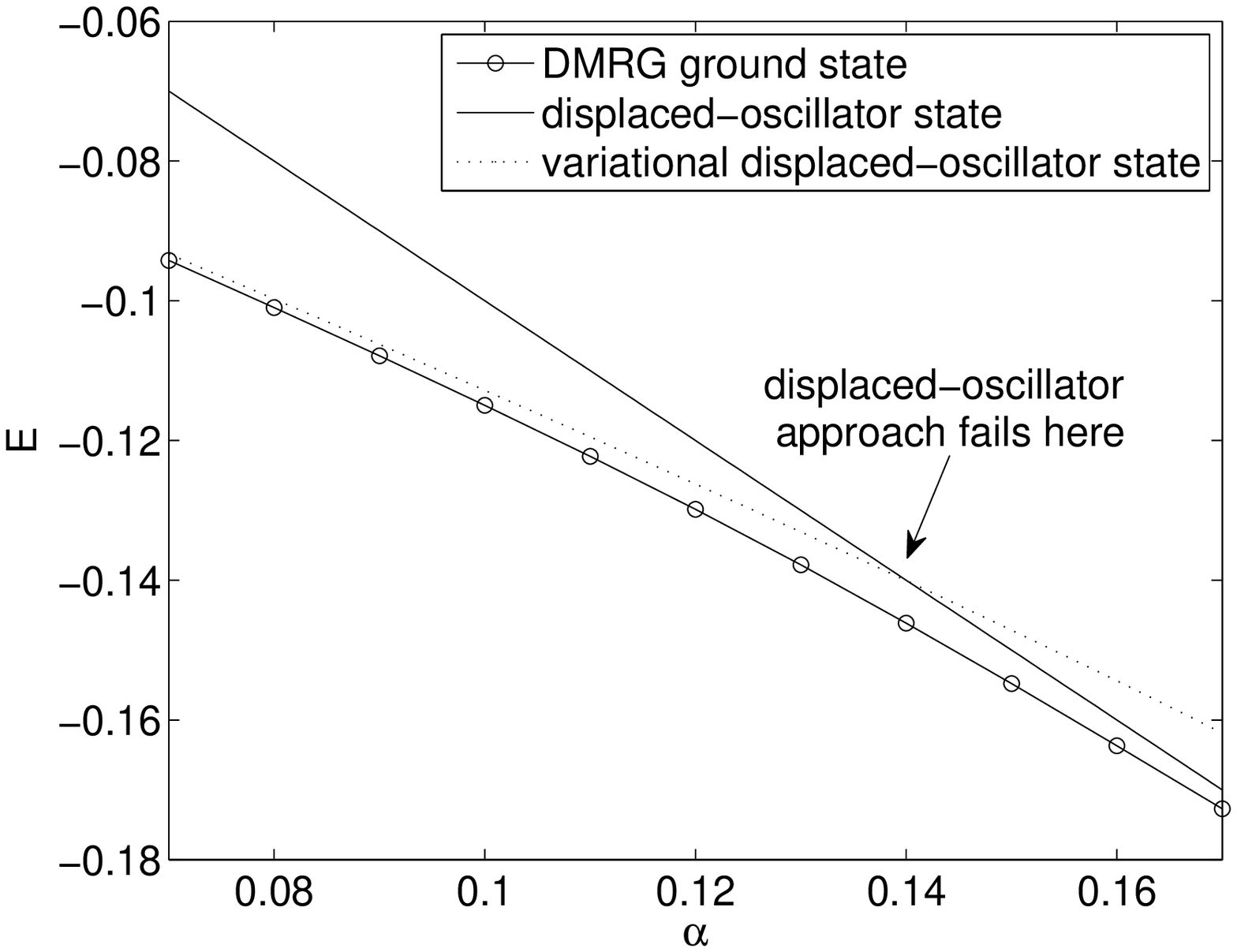}
        \caption{This figure shows the ground state energy
        comparison of the DMRG and variational approach for
        $\epsilon=0, s=0.5$, and $\Delta=0.1$.
        DMRG parameters are $N=201,N_b=30,M=20$, and $N_s=2$.}\label{fig7}
    \end{minipage}
\end{figure*}

Using the extrapolation scheme, the phase boundary for the
delocalized-localized transition of the spin-boson model for
$\epsilon=0$ and $\Delta=0.1$ is shown in Fig.\ref{fig6}, where
the result by NRG is also shown for comparison. It can be found
that the DMRG data are consistent with the NRG data quite well. It
also shows that the NRG data are always larger than the DMRG data.
Indeed, however, the results of NRG can be extrapolated to
thermodynamics limit, namely, one takes the NRG discretization
parameter $\Lambda\rightarrow1$, and smaller critical couplings
can be obtained.\cite{Bulla} In other words, both NRG and DMRG
show that the critical couplings before extrapolating to
thermodynamics limit are always larger than the true critical
couplings. This implies that the error of discretization on
determining the critical coupling is to lower the true $\alpha_c$,
but not to heighten the $\alpha_c$ as claimed in Ref.\
\onlinecite{Lv}.
In addition, the inset of Fig.\ \ref{fig6} also assures that our
DMRG calculations for Ohmic case are consistent with the NRG
result\cite{Bulla} and the well-known renormalization group
result,\cite{Leggett} i.e.,
$\alpha_c=1+\mathcal{O}(\Delta/\omega_c)$.

\section{Conclusion}
At the end of this paper, we want to compare the ground state
energy obtained by DMRG with the variational ground state energy
obtained by displaced-oscillator approach. As we mentioned in the
Introduction, the variational ground state will fail at a certain
point since the variational ground state is not yet
stable.\cite{Wong,Chen} The comparison of the ground state energy
is shown in Fig.\ \ref{fig7}. It shows that the DMRG ground state
energies are always lower than that of the variational
calculations. It is also clear that the DMRG ground state energy
approaches to variational ground state energy for small $\alpha$
and displaced-oscillator ground state energy for large $\alpha$,
and no discontinuous effective tunneling splitting and ground
state energy are observed by the DMRG calculations.

In conclusion, we have proposed a finite system DMRG algorithm to
deal with the spin-boson model. This algorithm is much more powerful
than the preceding study of this topic because it can treat more
than a thousand phonon modes.\cite{Nishiyama} In fact, we have tried
to calculate $10^4$ and $10^5$ phonon modes. Unfortunately, our
$32$-bit system is unable to tackle phonon modes on the $10^5$
magnitude due to out of memory. This difficultly, of course, can be
resolved in $64$-bit system or storing the data in hard disk.
Moreover, we obtain the critical couplings $\alpha_c$ by
extrapolating the pseudo-critical couplings to thermodynamic limit.
The phase diagram is compared with NRG and shows good agreement in
both sub-Ohmic and Ohmic cases.

\begin{acknowledgements}
We thank R. Bulla and E. Jeckelmann for helpful conversations, N.-H.
Tong for supplying the phase transition data obtained by NRG, X.-Q.
Wang for stimulating discussion. The iterative diagonalization is
done by the MATLAB program ``irbleigs."\cite{Baglama} This work was
supported by a grant from the Natural Science Foundation of China
under Grant No. 10575045.
\end{acknowledgements}


\begin{thebibliography}{99}
\bibitem {White} S. R. White, Phys. Rev. Lett. \textbf{69}, 2863
(1992); Phys. Rev. B \textbf{48}, 10305 (1993).
\bibitem {Schollwoeck} U. Schollw{\"o}ck, Rev. Mod. Phys. \textbf{77},
259 (2005).
\bibitem {Zhang} C. Zhang, E. Jeckelmann, and S. R. White, Phys. Rev.
Lett. \textbf{80}, 2661 (1998).
\bibitem {Weisse} A. Wei{\ss}e, H. Fehske, G. Wellein, and A. R. Bishop,
Phys. Rev. B \textbf{62}, R747 (2000).
\bibitem {Friedman} B. Friedman, Phys. Rev. B \textbf{61}, 6701
(2000).
\bibitem {Nishiyama} Y. Nishiyama, Eur. Phys. J. B \textbf{12}, 547
(1999).
\bibitem {Leggett} A. Leggett, S. Chakravarty, A. Dorsey, M.
Fisher, A. Garg, and W. Zwerger, Rev. Mod. Phys. \textbf{59}, 1
(1987).
\bibitem {Weiss} U. Weiss, {\it Quantum Dissipative Systems} (World Scientific, Singapore,
1999).
\bibitem {sp}H. Spohn and R. D{\"u}mcke, J. Stat. Phys. {\bf 41}, 389 (1985).
\bibitem {keh}S. K. Kehrein and A. Mielke, Phys. Lett. A {\bf 219}, 313 (1996).
\bibitem {Silbey} R. Silbey and R. Harris, J. Chem. Phys. \textbf{80},
2615 (1984).
\bibitem {Chin} A. Chin and M. Turlakov, Phys. Rev. B \textbf{73},
75311 (2006).
\bibitem {Wong} H. Wong and Z.-D. Chen, Phys. Rev. B \textbf{76}, 77301(2007).
\bibitem {Lv} Z. L{\"u} and H. Zheng, Phys. Rev. B
\textbf{75}, 54302 (2007).
\bibitem {Bulla} R. Bulla, N.-H Tong, and M. Vojta, Phys. Rev. Lett.
\textbf{91}, 17061 (2003); R. Bulla, H.-J. Lee, N.-H. Tong, and M.
Vojta, Phys. Rev. B \textbf{71}, 45122 (2005).
\bibitem {Li} M.-R. Li, K. Le Hur, and W. Hofstetter, Phys. Rev.
Lett. \textbf{95}, 86406 (2005).
\bibitem {qpt}S. Sachdev, {\it Quantum Phase Transition}, (Cambridge
University Press,  Cambridge, England, 1999).
\bibitem {qp}M. Vojta, Rep. Prog. Phys. {\bf 66}, 2069 (2003).
\bibitem {Chen} Z.-D. Chen and H. Wong, arXiv:0705.1670 (unpublished).
\bibitem {Legeza} {\"O}. Legeza and G. F{\'a}th, Phys. Rev. B
\textbf{53}, 14349 (1996).
\bibitem {note1} This is rigorous for targeting pure state. However, for
simplicity, we can keep two states here even though we target more
than one states because the error can be reduced in the following
sweeping process.
\bibitem {White2} S. R. White, Phys. Rev. B \textbf{72}, R180403
(2005).
\bibitem {White3} S. R. White, Phys. Rev. Lett. \textbf{77}, 3633
(1996).
\bibitem {Bulla2} R. Bulla, T. Costi, and T. Pruschke, Rev. Mod.
Phys. (in press), see also arXiv:cond-mat/0701105.
\bibitem {Anders} F. B. Anders, R. Bulla, M. Vojta, Phys. Rev. Lett.
\textbf{98}, 210402 (2007).
\bibitem {Hur} K. Le Hur, P. Doucet-Beaupr{\'e}, and W. Hofstetter,
Phys. Rev. Lett. \textbf{99}, 126801 (2007).
\bibitem {Boschi} C. Degli Esposti Boschi and F. Ortolani, Eur.
Phys. J. B \textbf{41}, 503 (2004).
\bibitem {Capone} M. Capone, S. Caprara, Phys. Rev. B \textbf{64},
184418 (2001).
\bibitem {Juo} A. Juozapavi{\v c} ius, S. Caprara, A. Rosengren,
Phys. Rev. B \textbf{56}, 11097 (1997).
\bibitem {Juo2} A. Juozapavi{\v c}ius, L. Urba, S. Caprara, and A.
Rosengren, Phys. Rev. B \textbf{60}, 14771 (1999).
\bibitem {Baglama} J. Baglama, D. Calvetti, and L. Reichel, ACM T. Math.
Software \textbf{29}, 337 (2003).
\end{thebibliography}
\end{document}